\documentclass[pdflatex,sn-mathphys-num]{sn-jnl}

\usepackage{graphicx}%
\usepackage{multirow}%
\usepackage{amsmath,amssymb,amsfonts}%
\usepackage{amsthm}%
\usepackage{mathrsfs}%
\usepackage[title]{appendix}%
\usepackage{xcolor}%
\usepackage{textcomp}%
\usepackage{manyfoot}%
\usepackage{booktabs}%
\usepackage{algorithm}%
\usepackage{algorithmicx}%
\usepackage{algpseudocode}%
\usepackage{listings}%
\usepackage{colortbl}
\usepackage[table]{xcolor} 

\theoremstyle{thmstyleone}%
\theoremstyle{thmstyletwo}%

\theoremstyle{thmstylethree}%

\begin{document}

\title[Article Title]{Phenome-Wide Multi-Omics Integration Uncovers Distinct Archetypes of Human Aging}


\author[]{\fnm{Huifa} \sur{Li}}

\author[]{\fnm{Feilong}\sur{Tang}}
\author[]{\fnm{Haochen} \sur{Xue}}
\author[]{\fnm{Yulong} \sur{Li}}
\author[]{\fnm{Xinlin}\sur{Zhuang}}
\author[]{\fnm{Bin  } \sur{Zhang}} 
\author[]{\fnm{Eran  } \sur{Segal}}
\author[]{\fnm{Imran  } \sur{Razzak}}

\affil[]{\orgname{Mohamed bin Zayed University of Artificial Intelligence}, \orgaddress{\city{Abu Dhabi}} \\
\{Huifa.L,Feilong.Tan, Haochen.Xue, Yulong.Li, Xinlin.Zhuang, Bin.Zhang, Eran.Segal, Imran.Razzak\}@mbzuai.ac.ae
}


\abstract{
Aging is a highly complex and heterogeneous process that progresses at different rates across individuals, making biological age (BA) a more accurate indicator of physiological decline than chronological age. While previous studies have built aging clocks using single-omics data, they often fail to capture the full molecular complexity of human aging. In this work, we leveraged the Human Phenotype Project\footnote{\url{https://knowledgebase.pheno.ai/}}, a large-scale cohort of 10,000 adults aged 40–70 years, with extensive longitudinal profiling that includes clinical, behavioral, environmental, and multi-omics datasets—spanning transcriptomics, lipidomics, metabolomics, and the microbiome. By employing advanced machine learning frameworks capable of modeling nonlinear biological dynamics, we developed and rigorously validated a multi-omics aging clock that robustly predicts diverse health outcomes and future disease risk. Unsupervised clustering of the integrated molecular profiles from multi-omics uncovered distinct biological subtypes of aging, revealing striking heterogeneity in aging trajectories and pinpointing pathway-specific alterations associated with different aging patterns. These findings demonstrate the power of multi-omics integration to decode the molecular landscape of aging and lay the groundwork for personalized healthspan monitoring and precision strategies to prevent age-related diseases. 

}




\maketitle

\section{Introduction}\label{sec1}
In the life sciences, age serves as a fundamental metric for gauging the passage of time from birth to a given point in life \cite{niccoli2012ageing,partridge2018facing}. Behind this temporal measure lies a complex biological reality. Aging is a natural process characterized by the progressive decline of physiological functions, driven by the gradual accumulation of molecular and cellular damage over time. This systemic deterioration is far from benign—aging is the primary risk factor for a wide range of chronic diseases, including cancer, cardiovascular disorders, and neurodegenerative conditions \cite{partridge2018facing,chang2019measuring,laconi2020cancer,montegut2024aging,guo2022aging}. 

While chronological age (CA) provides the basic framework for understanding the aging process, it serves only a coarse proxy for an highly variable physiological phenomenon. Yet, this framework fails to account for the substantial interindividual heterogeneity in health trajectories and disease susceptibility. In contrast, biological age (BA) aims to quantify the rate and state of physiological decline resulting from the accumulation of molecular and cellular damage and more accurately reflects a patient's physiological state and resilience than chronological age \cite{ikram2024use,jackson2003biological,levine2013modeling,salih2023conceptual}. By doing so,it offers accurate reflection of a patient’s physiological state and resilience than chronological age, enabling improved risk prediction, disease prevention, and personalized care decisions \cite{rutledge2022measuring}.

The marked heterogeneity in aging rates among individuals is not only a central determinant of morbidity and mortality risk, but also underlies the asynchronous decline of distinct physiological systems \cite{yu2025cross}. Therefore, accurately quantifying biological age is essential for capturing and understanding this age-related, individualized risk \cite{langenberg2023biological}. Biological age is a dynamic, multidimensional construct that integrates the cumulative influences of genetic background, environmental exposures, and lifestyle factors, enabling the differentiation of health status among individuals with the same chronological age. Aging clocks are biomarker-based models that estimate physiological age by applying statistical or machine learning algorithms to high-dimensional molecular or physiological data \cite{han2024ticking}. These clocks are designed to capture and quantify the rate of aging, which varies significantly across individuals. The selection of biomarkers and modeling approaches directly impacts the predictive performance, biological interpretability, and ultimate clinical translatability of aging clocks \cite{moqri2024validation}.

Omics-based aging clocks have emerged as powerful tools capable of revealing an individual's aging status and disease risk with unprecedented precision. Leveraging high-throughput omics technologies, researchers have developed a range of biological clocks to quantify the complex process of aging. Among these, epigenetic clocks based on DNA methylation represent a pioneering effort—not only can they accurately predict chronological age, but they have also been widely validated as robust predictors of all-cause mortality and healthspan \cite{horvath2018dna,tong2024quantifying}. Building on this foundation, researchers have explored additional molecular layers of aging biomarkers. At the transcriptomic level, one of the first RNA-based aging clocks was developed in 2015 using whole-blood RNA sequencing data \cite{peters2015transcriptional}. Given that proteins serve as the direct effectors of cellular function, proteomic clocks based on plasma proteins have shown particularly strong clinical potential in predicting various age-related diseases and mortality risk \cite{argentieri2024proteomic}. Beyond host-derived molecular signals, recent attention has turned toward the microbiome—our symbiotic microbial communities—which act as dynamic sensors linking the host with the external environment and profoundly reflect or even influence aging trajectories through immune and metabolic regulation \cite{galkin2020human,huang2020human}. Emerging evidence suggests that a microbiome age that deviates from a healthy reference is significantly associated with increased mortality, reduced physical function, and heightened systemic inflammation. Despite these advances, there remains a critical gap: no study to date has systematically integrated multidimensional omics data within a large-scale prospective cohort to develop and validate a multi-omics aging clock designed to predict the risk of diverse disease outcomes.

Beyond biomarker selection, a central element in constructing aging clocks lies in the underlying statistical framework. A growing consensus holds that human aging is not a uniform, linear process. Compelling evidence comes from clinical epidemiology: the incidence of major age-related diseases does not increase linearly with age, but instead follows an accelerated, near-exponential trajectory. For example, in the U.S. population, the prevalence of cardiovascular disease rises from approximately 40\% in individuals aged 40–59 to around 75\% at ages 60–79, and reaches 86\% in those over 80 \cite{rodgers2019cardiovascular}. This macroscopic clinical pattern reflects underlying nonlinear molecular dynamics at the microscopic level. Numerous studies have documented that during aging, gene expression, protein abundance, and metabolite levels exhibit complex nonlinear patterns \cite{shen2024nonlinear,lehallier2019undulating}. As such, linear models—which assume a simple additive relationship between biomarkers and aging—are inherently limited in their ability to capture and interpret this complexity. These models may overlook critical inflection points, threshold effects, or synergistic and antagonistic interactions between biological pathways that define the trajectory of aging. In contrast, nonlinear statistical frameworks—such as ensemble learning models and deep neural networks—are inherently more capable of modeling the intricate data structures and dynamic behaviors that characterize the biological aging process. These methods hold promise for capturing the true, multidimensional trajectory of human aging with greater fidelity.

Here we leveraged a large-scale, well-characterized human cohort (Human-Phenotype Project, HPP) \cite{shilo202110} consisting of 10K longitudinal study, which includes comprehensive clinical, physiological, behavioral, environmental, and multi-omics profiling, to investigate the complex dynamics of human aging. 
Most existing aging clocks (like epigenetic clocks) produce one output. By integrating high-dimensional data spanning the transcriptome, lipidome, metabolome, and microbiome, we developed and rigorously validated a novel multi-omics aging clock and provided holistic and biologically informative picture of systemic decline such as risk association. 
Through unsupervised clustering of the integrated omics landscape, we identified distinct molecular subtypes, thereby uncovering the profound molecular and phenotypic heterogeneity inherent to the human aging process. 
Subsequent pathway enrichment and network analyses of these subtypes enabled us to deconstruct the biological consequences of aging and demonstrate how key biological processes are dynamically reweighted across divergent aging trajectories.

\begin{figure*}[htbp]
    \centering
    \includegraphics[width=0.95\linewidth]{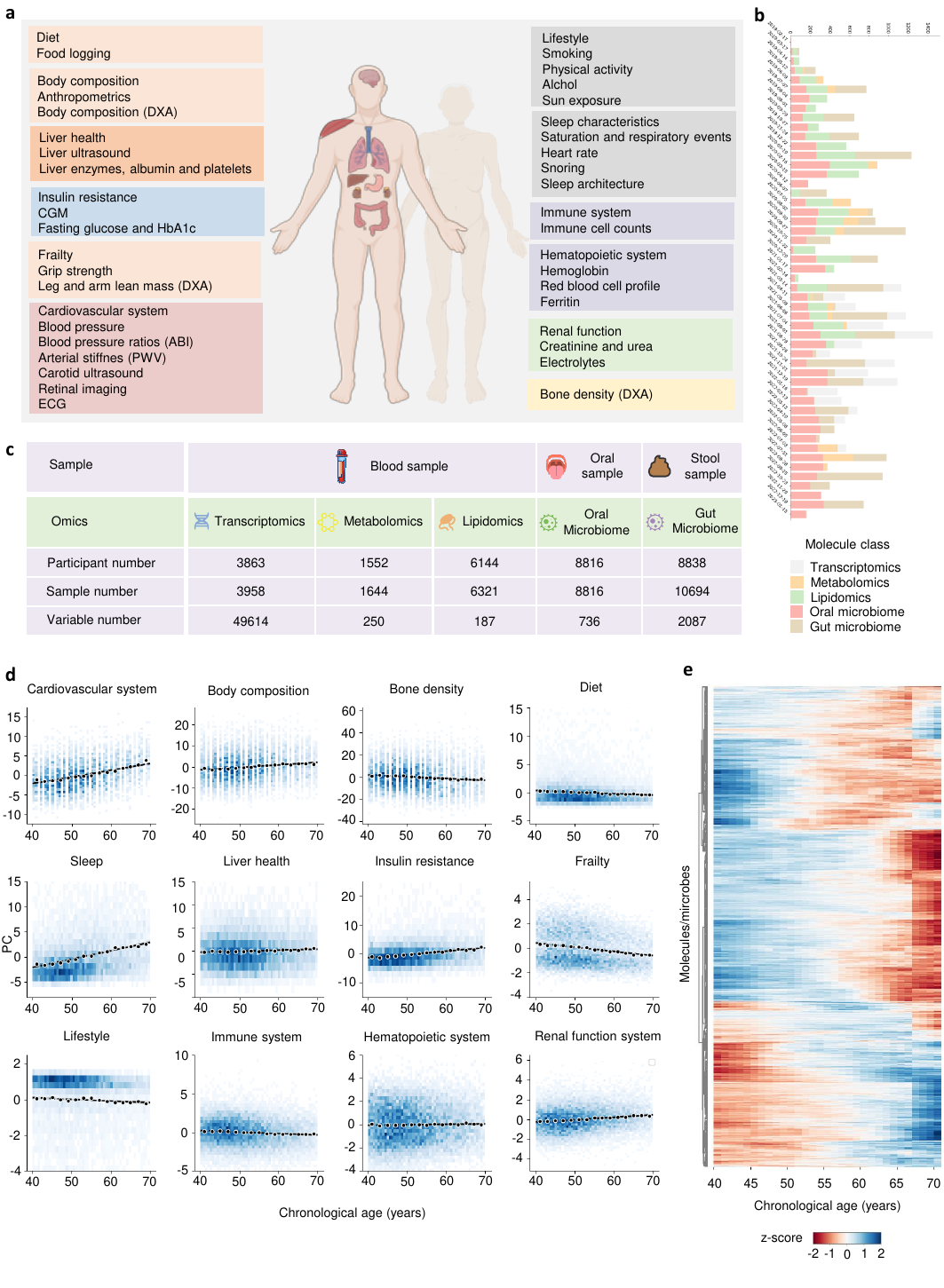}
    \caption{\textbf{Comprehensive data collected in an aging cohort.} 
    \textbf{a}, Illustration of the comprehensive clinical, physiological and behavioral data collected in the 10K study which are assigned into 12 system-level categories. 
    \textbf{b}, Collection time range and sample numbers for the cohort.
    \textbf{c}, Sample collection and multi-omics data acquisition of the cohort. Three types of biological samples were collected, and 5 types of omics data were acquired. 
    \textbf{d}, Spearman correlation between the first principal component and chronological ages for each type of systems.
    \textbf{e}, Heatmap depicting the dynamic changing molecules and microbes during human aging.}
    \label{fig:human-system}
\end{figure*}

\section{Results}\label{sec2}
\subsection{System-wide characterization in a large healthy aging cohort}
To explore the phenotypes of various systems during aging, we analyzed an ongoing collection project HPP, which is a large-scale phenotyped cohort consisting of ethnically diverse people in Israel, most of whom are healthy and have received higher education.
This deep phenotyped cohort includes more than 10,000 participants aged 40-70 years with a 5-year follow-up and consist of not only multi-omics data but also clinical, physiological, and behavioral data, which can be divided into 12 body systems (Fig. \ref{fig:human-system}a, b).
The BMI of the participants ranged from 15.73 to 47.23. Among the participants, 47.49\% were male. For each follow-up, the blood, stool, and oral swab samples were collected. A total of 31,433 biological samples (including 11,923 blood samples, 10,694 stool samples, and 8,816 oral swab samples) were collected (Fig. \ref{fig:human-system}c). 
The biological samples were used for multi-omics profiling (including transcriptome from peripheral blood mononuclear cells, metabolomics in plasma, lipidomics in plasma, gut microbiome, and oral microbiome). A total of 52,874 biological features (including 49,614 transcripts, 250 metabolites, 187 lipids, 736 gut microbiome taxa, and 2,087 oral microbiome taxa) were collected, generating 9,787,326,308,545 data points.

To elucidate the multi-system effects of aging, we examined the association between HPP data and physiological age. Most systemic indicators displayed significant age dependency (Fig. \ref{fig:human-system}d), with cardiovascular and sleep measures showing the strongest positive correlations with age. These findings reinforce the view that aging is a complex, multisystem process and underscore the utility of our HPP dataset for quantifying and tracking human aging trajectories. Moreover, biological features exhibited pronounced nonlinear changes across age (Fig. \ref{fig:human-system}e). Together, this extensive multi-omics resource enables comprehensive investigation of molecular, phenotypic, and microbial dynamics during aging. The large cohort (12,000 participants) provides exceptional power to detect intricate aging-related changes across both molecular and functional dimensions.

\subsection{Multi-omics aging clocks predict biological age}
\begin{figure*}[!t]
    \centering
    \includegraphics[width=1.0\linewidth]{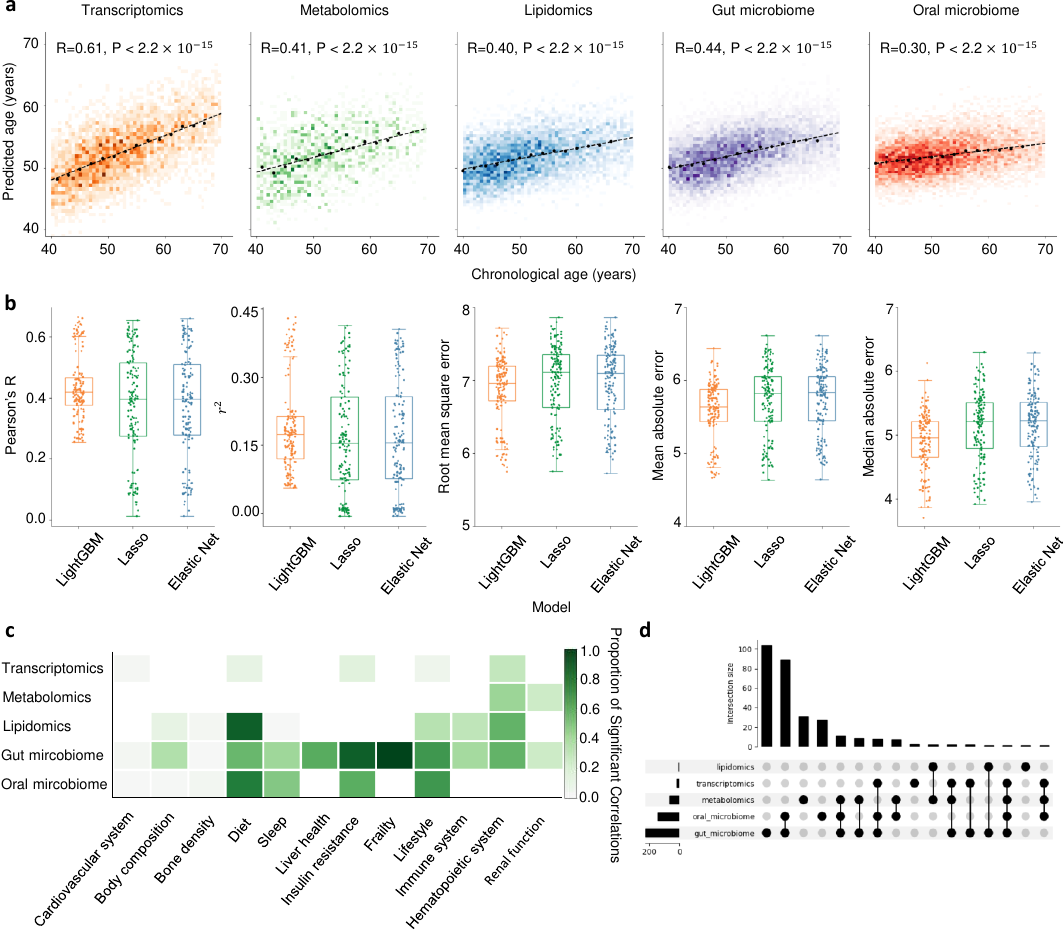}
    \caption{\textbf{Internal validation demonstrates the accuracy, robustness, and systemic relevance of multi-omics aging clocks.} \textbf{a}, Pearson correlation (two-sided test) is plotted between actual chronological age and predicted age for each omics type. Linear fit is plotted as a dashed line.  \textbf{b}, To verify the robustness of the aging clock models, the training set and internal validation set are randomly resampled, and the model is trained for 10 times. Pearson correlation (two-sided test), $r^2$, r.m.s.e., m.a.e. and med.a.e are used as evaluation criteria. \textbf{c}, Proportion of biomarkers significantly associated with each omics AgeAccel across different physiological systems. For each system, elements represent the percentage of all measured biomarkers within that system that are significantly correlated (FDR-adjusted p $<$ 0.05, Pearson correlation) with a given omics AgeAccel. This analysis highlights the systemic breadth of each clock. \textbf{d}, UpSet plot showing the number of unique and shared biomarkers significantly associated with the different aging omics AgeAccels. The horizontal bars indicate the total number of significant biomarkers for each omics AgeAccel. The upper vertical bars show the number of biomarkers in each intersection, with the matrix below indicating which omics AgeAccels are included in that intersection. This illustrates the specificity and overlap of biological signals captured by each type of omics AgeAccels.}
    \label{fig:aging_clock}
\end{figure*}

We developed a multi-omics aging clock using five types of omics data derived from human PBMC, feces, and saliva samples.
During the training phase, we compared three machine learning methods (least absolute shrinkage selection operator (LASSO) \cite{tibshirani1996regression}, Elastic Net \cite{ho1995random}, LightGBM \cite{ke2017lightgbm}) to train multi-omics age clock models.
We evaluated the performance of the aging clock by comparing the predicted age with the CA (Pearson’s R, $r^2$, root mean square error (r.m.s.e.), median absolute error (m.a.e.) and median absolute error (med.a.e)). A cross-validation step was implemented, where 80\% of the samples were randomly selected for model training, while the remaining 20\% of the data were used as an internal validation set. Fig. \ref{fig:aging_clock}a underscored the accuracy of our aging clocks.
In our initial experiments, we found that LightGBM showed superior performance across all omics types. Consequently, we selected it as the final model and used SHapley Additive exPlanations (SHAP) \cite{lundberg2020local} to identify omics features associated with predicting chronological age.
The strong correlation between chronological age and predicted transcriptional age demonstrated the accuracy of our aging clock.
We further confirmed the robustness of the aging clock model through 10-fold cross-validation, which showed only slight performance changes in all metrics (Fig. \ref{fig:aging_clock}b).

To evaluate the physiological relevance of age acceleration across biological strata, we assessed associations between five omics-based age-acceleration (AgeAccel) scores and a comprehensive panel of parameters spanning 12 physiological and lifestyle domains (Fig. \ref{fig:aging_clock}c). AgeAccel scores, computed as residuals from regressing biological age on chronological age, differed substantially in their links to systemic health, after adjusting for multiple hypotheses (false discovery rate (FDR) $<$ 0.05). 
Of these, diet, lifestyle and hematopoietic system were significantly associated with four omics-based AgeAccel metrics. 
Gut-microbiome AgeAccel showed the broadest association profile (significant associations with diet, liver health, insulin resistance, frailty, lifestyle and hematopoietic system) and was the only omics-derived AgeAccel associated with frailty, consistent with previous findings by \cite{pu2024gut}. 
Oral-microbiome AgeAccel showed a similar but weaker pattern. The oral-microbiome AgeAccel showed the strongest association with diet ($prop$=0.78), independent of sex and BMI. This pattern aligns with previous reported links whereby diet shapes the oral microbiota by providing nutrients and selecting organisms adapted to specific dietary resources \cite{sedghi2021oral,adler2016diet}. 
In contrast to the broad profiles of the microbiome-based metrics, the remaining molecular AgeAccel clocks exhibited more circumscribed physiological associations. The Transcriptomic-AgeAccel was predominantly linked to immune and inflammatory processes, with biomarkers of the hematopoietic system accounting for about one-third of its significant associations. 
Weaker, yet significant, links were also observed with diet ($prop$=0.11), insulin resistance ($prop$=0.15), and lifestyle ($prop$=0.06), while its connection to the cardiovascular system was minimal ($prop$=0.02). 
The Metabolomic-AgeAccel displayed an even more specific signature, being almost exclusively associated with the hematopoietic system ($prop$=0.43). 
Conversely, the lipidomic-AgeAccel captured a more pleiotropic signal; it was most strongly dominated by associations with diet ($prop$=0.89) but was also substantially linked to the lifestyle ($prop$=0.33) and immune ($prop$=0.30) systems in addition to hematopoietic system ($prop$=0.57).

We quantified specificity and redundancy across AgeAccel metrics by analyzing set intersections of FDR-significant associations (Fig. \ref{fig:aging_clock}d). 
After controlling for multiple testing, each AgeAccel retained a non-overlapping subset of parameters, with the gut-microbiome AgeAccel showing the largest unique subset (104 of 227 parameters), consistent with prior evidence that the gut microbiome captures aging-related and systemic signals \cite{bosco2021aging,shuai2022mapping}. 
Beyond unique signals, we detected non-random overlaps: the largest intersection involved parameters associated with both gut and oral-microbiome AgeAccel, in line with microbiome-based aging clocks and reports on the oral–gut axis \cite{ray2020oral,kunath2024oral}. 
Furthermore, a smaller but crucial set of biomarkers was significantly associated with four aging AgeAccels, suggesting a conserved, multi-system aging signal that is detectable across most biological layers of organization.


\subsection{Systemic Signatures and Molecular Dynamics of Aging Phenotypes}
\begin{figure*}[htbp]
    \centering
    \includegraphics[width=1.0\linewidth]{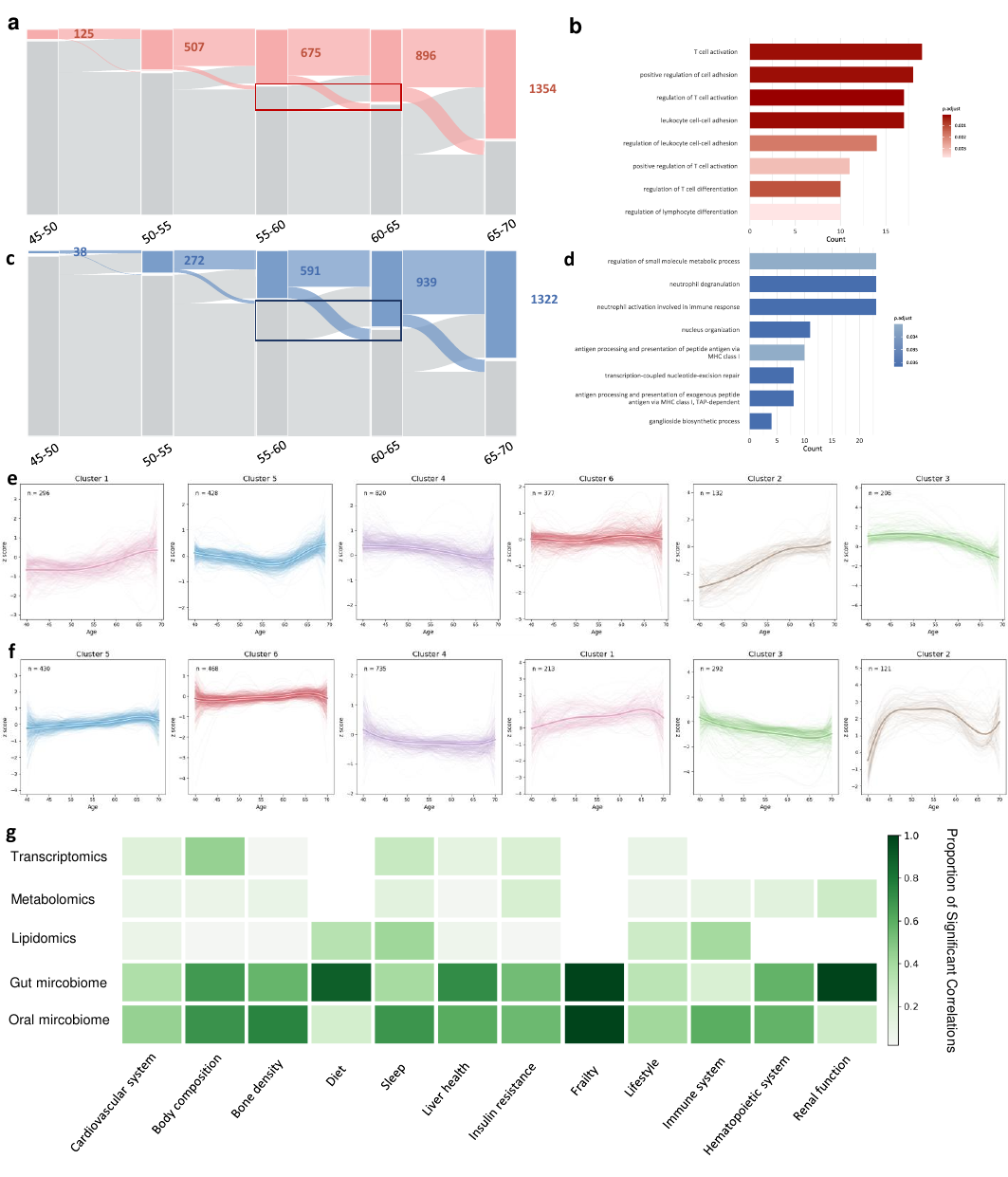}
    \caption{\textbf{Distinct molecular aging trajectories and systemic signatures in accelerated and decelerated aging.}
    \textbf{a-d}, Dynamics of significantly altered biological features across age in the accelerated (\textbf{a, b}) and decelerated (\textbf{c, d}) aging groups.
    \textbf{a, c}, Sankey diagrams showing the number of features that significantly increase or decrease with age relative to the 40–45-year baseline (two-sided Mann–Whitney U-test, FDR-adjusted P$<$0.05).
    \textbf{b, d}, Pathway enrichment of genes significantly altered between the 55–60 and 60–65-year groups.
    \textbf{e, f}, Temporal trajectory clusters of all multi-omics features for the accelerated (\textbf{e}) and decelerated (\textbf{f}) aging groups, identified using Fuzzy C-means clustering. Thin lines represent individual z-scored trajectories, and thick lines denote cluster centroids. The number of features (n) per cluster is indicated.
    \textbf{g}, Heatmap showing the percentage of features that are significantly different between the accelerated and decelerated aging groups across various physiological systems. Color intensity corresponds to the percentage of significant features within each system, identified by one-way ANOVA.
    }
    \label{fig:clustering}
\end{figure*}

To understand the biological processes that distinguish accelerated from decelerated aging, we first classified participants into aging-accelerated and aging-decelerated groups based on AgeAccel scores, and further categorized them into distinct age stages according to their ages and investigated the altered biological features within each age stage compared to the baseline (40–45 years old; Methods). 
Our analysis revealed that accelerated aging is marked by an early and large-scale disruption of molecular and microbial homeostasis. A substantial number of features showed significant changes from the 40–45 age baseline as early as the 50–55 age bracket, with these alterations accumulating progressively into later life (Fig. \ref{fig:clustering}a). 
We then focused on features that were not significantly altered at 55–60 years but became significant at 60–65 years, and performed differential expression and Gene Ontology (GO) enrichment analyses \cite{ashburner2000gene} for these late-changing features in both the accelerated and decelerated aging groups (Methods). For instance, in accelerated aging, enrichment for T-cell activation \cite{mittelbrunn2021hallmarks,goronzy2019mechanisms} indicates an adaptive-immune, adhesion-driven inflammatory state that is consistent with chronic antigenic stimulation and tissue infiltration (Fig. \ref{fig:clustering}b).
In stark contrast, the decelerated aging group exhibited a blunted and delayed molecular aging phenotype, with a significantly smaller number of features changing in early mid-life while the number of features that change during the transition from middle age to old age increases dramatically (Fig. \ref{fig:clustering}c). 
Unlike accelerated aging, decelerated aging is marked by enrichment of regulation of small-molecule metabolic process, aligning with preserved metabolic and redox homeostasis that supports genome maintenance and antigen presentation while restraining excessive immune activation\cite{qin2024enhancing,lu2023cellular} (Fig. \ref{fig:clustering}d).

To identify the dominant temporal patterns driving these differences, we clustered the trajectories of all molecular features using an unsupervised fuzzy c-means clustering approach \cite{shen2024multi}. This revealed fundamentally different aging archetypes between the groups (Fig. \ref{fig:clustering}e, f). The accelerated aging cohort was characterized by trajectories showing precipitous changes during mid-life. For instance, a large cluster of features (Fig. \ref{fig:clustering}e, Cluster 4, n=820) maintained stability until approximately age 55, after which they entered a steep and continuous decline. Another cluster (Fig. \ref{fig:clustering}e, Cluster 5, n=428) showed a sharp increase during the same period. Conversely, the trajectories in the decelerated aging group were primarily defined by prolonged stability and more gradual, later-life changes. The corresponding declining cluster (Fig. \ref{fig:clustering}f, Cluster 4, n=735) was smaller and showed a much gentler slope beginning later in life, while other dominant clusters remained notably stable until after age 60 (Fig. \ref{fig:clustering}e). However, cluster 2 (Fig. \ref{fig:clustering}f, n=121) increases rapidly before the age of 45, remains stable, and then decreases rapidly after the age of 60.

We next investigated which physiological systems were most different between the two aging groups. The divergence was not uniform across biological layers. Instead, the most pronounced differences were concentrated in the gut and oral microbiomes (Fig. \ref{fig:clustering}g). A high proportion of microbial features were significantly different between the two groups, particularly those associated with diet, frailty, and lifestyle. This indicates that the distinction between accelerated and decelerated aging is strongly linked to a microbiome-centric axis.

Taken together, these analyses demonstrate that accelerated aging is not simply a faster version of the typical aging process but a distinct phenomenon characterized by an early, microbiome-associated dysregulation that precedes sharp, mid-life shifts in the trajectories of specific molecular families.

\subsection{Uncovering waves of aging-related molecules during aging}
\begin{figure*}[htbp]
    \centering
    \includegraphics[width=1.0\linewidth]{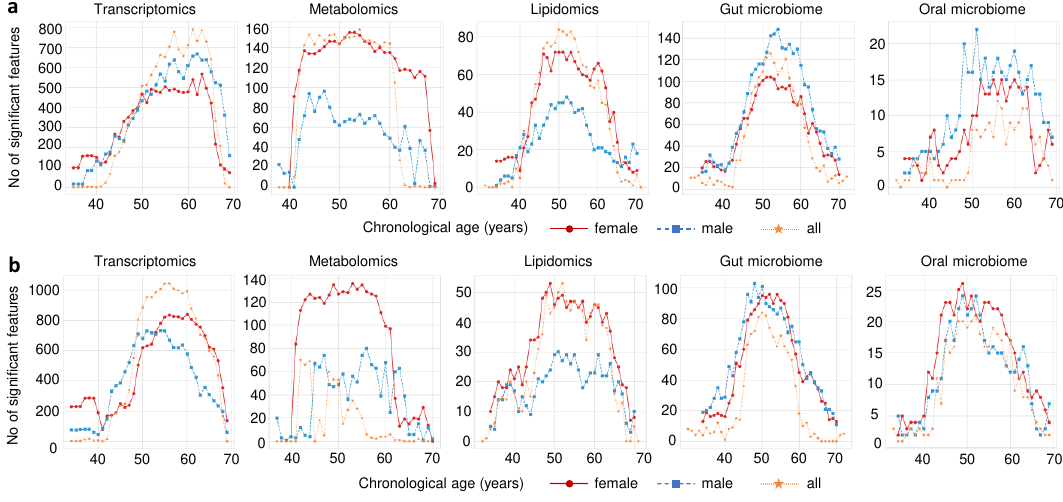}
    \caption{\textbf{Accelerated and decelerated aging exhibit distinct temporal waves of molecular change across multi-omics layers.}
    \textbf{a}, The number of significantly age-associated biological features across chronological age in the accelerated aging group. Each panel corresponds to a specific omics type. The y-axis represents the count of features significantly correlated with age (FDR-adjusted q $<$ 0.05) at each age point. Analyses are presented for the entire cohort (all, orange), females only (red), and males only (blue).
    \textbf{b}, Same as \textbf{a}, but for the decelerated aging group. The plots reveal different magnitudes and temporal peaks of significant molecular changes compared to the accelerated aging group, particularly highlighting sex-specific differences in aging dynamics.}
    \label{fig:functional_analysis}
\end{figure*}

To capture age-specific dynamics, we applied a modified DE-SWAN \cite{lehallier2019undulating}, which tests sliding 20-year windows in 10-year parcels across the lifespan and tabulates significant molecules and microbes (q $<$ 0.05) separately for accelerated and decelerated aging, with sex-stratified analyses (Fig. \ref{fig:functional_analysis}a, b). Across all five omics layers we observed robust age-related waves with two consistent crests—one around 45–55 years and a second near 60–65 years—followed by a decline after 67–70 years. Transcriptomics and lipidomics showed the largest amplitudes, metabolomics exhibited a sustained mid-life plateau, and both gut and oral microbiomes displayed concordant crests. Compared with the decelerated mode, the accelerated mode showed a sharper peak, but had an earlier peak in the oral microbiome; the decelerated mode exhibited later and broader crests, particularly in transcriptomes. Sex-stratified curves revealed higher crests in females for metabolomics and lipidomics, whereas males showed a higher mid-life crest in the oral microbiome and a slightly later crest in transcriptomes. The cross-layer consistency supports these crests as molecular milestones of aging and points to coordinated, system-level remodeling rather than single-layer changes.

\subsection{Multi-omics aging clocks predict multimorbidity}

\begin{figure*}[htbp]
    \centering
    \includegraphics[width=1.0\linewidth]{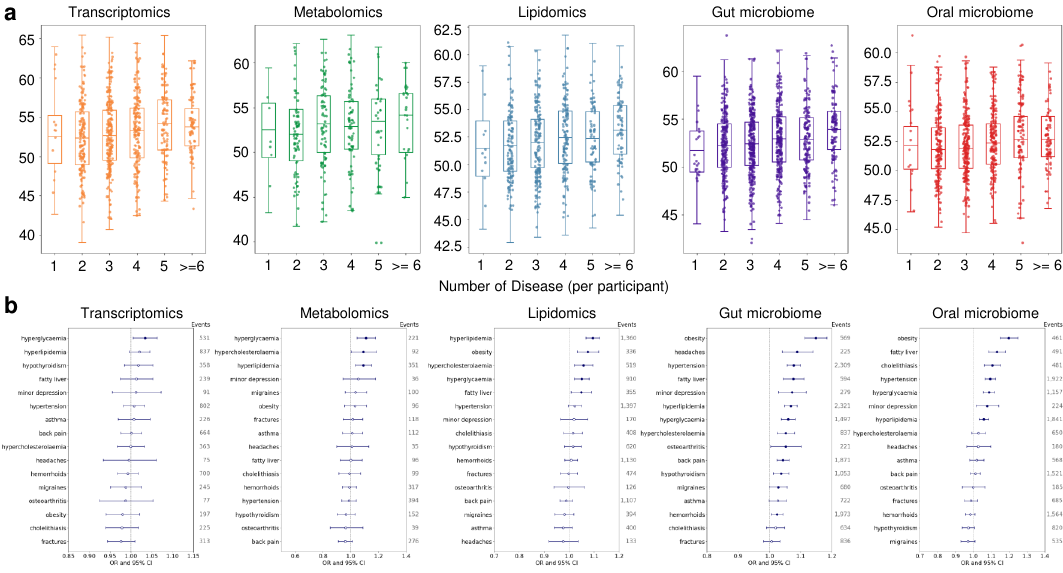}
    \caption{\textbf{Multi-omics aging clocks associated with age-related diseases.} 
    \textbf{a}, After adjusting for covariates, multi-omics aging clocks was significantly correlated with multimorbidity in the older population analyzed. 
    \textbf{b}, Associations between multi-omics aging clocks and mortality and disease incidence using OR.}
    \label{fig:disease}
\end{figure*}

To further understand the impact of age-related chronic inflammation on age-related pathologies, we calculated regression analyses between the total number of age-related diseases (multimorbidity) and the multi-omics aging clock. Multimorbidity is a global health research priority and has become the gold standard for aging research because it represents the accumulation of physiological damage in an individual\cite{skou2022multimorbidity,pearson2019multimorbidity,barnes2015mechanisms,marengoni2011aging,barnett2012epidemiology}. All of these disease features were binary. For these analyses, we controlled for sex, BMI, as each of these variables has reported effects in the etiology of age-related pathologies. We observed a significant correlation between the multi-omics aging clock and multimorbidity in the elderly in this study (Fig. \ref{fig:disease}a). This highlights the key role of the multi-omics aging clock in identifying the accumulation of physiological damage during aging.

Five different omics AgeAccel scores and the incidence of 16 common non-cancer diseases showed different specificities. Since our study has not yet been followed up for a long time, we examined whether the calculated AgeAccel scores were associated with medical diagnosis at baseline. 
We used a logistic regression model to estimate the association between the five AgeAccel scores and 16 common non-cancer diseases after adjusting for actual age, sex, and BMI. 
The analysis results revealed that the association between the five omics aging AgeAccel scores and diseases showed high specificity, among which the microbiome AgeAccel scores showed the most extensive and strongest association (Fig.  \ref{fig:disease}b). 
The gut microbiome AgeAccel score showed the strongest disease association signal among all AgeAccel scores, and was statistically significantly positively correlated with 13 of the 16 diseases. The strongest association was in obesity diseases. Only asthma, cholelithiasis and fractures did not show significant associations (Fig. \ref{fig:disease}b). 
The oral microbiome AgeAccel score also showed a similar but weaker range of associations, significantly associated with 7 of the 16 diseases. Unlike the gut microbiome, its association with cholelithiasis is statistically significant (Fig. \ref{fig:disease}b). 
Compared with the microbiome AgeAccel score, the molecular omics AgeAccel score derived from blood has a narrower range of associations, mainly concentrated in cardiometabolic diseases. The transcriptome AgeAccel score has the weakest association signal, with only a weak but statistically significant association with hyperglycemia (Fig. \ref{fig:disease}b).


\section{Discussion}\label{sec3}
In this study, we leveraged a large-scale, deeply phenotyped prospective cohort to develop and validate a multi-omics aging clock, providing a comprehensive framework for quantifying the complexity of human aging (Fig. \ref{fig:human-system}a). Our central finding is that human aging is not a uniform, linear process but a dynamic and heterogeneous phenomenon characterized by distinct molecular trajectories and sex-specific dynamics. By integrating data across the transcriptome, metabolome, lipidome, and microbiome, our work offers a systems-level view of aging, identifies robust biomarkers predictive of health status, and pinpoints the microbiome as a central hub in the aging process (Fig. \ref{fig:aging_clock}a).

Our multi-omics approach represents a significant advance over previous aging clocks developed from single data layers, such as DNA methylation. While each omics layer provides a unique window into the aging process, their integration offers a more holistic and biologically informative picture of systemic decline. A particularly striking finding was the superior performance and broader systemic relevance of microbiome-based clocks. Both the gut and oral microbiome clocks demonstrated the most extensive associations with diverse physiological systems, including diet, sleep, liver health, and immunity. This suggests the microbiome acts as a highly sensitive sensor, integrating signals from genetic, environmental, and lifestyle factors to reflect overall host health. While prior work has linked the microbiome to aging and age-related diseases , our findings position it as a powerful and broad-spectrum indicator of biological age and systemic health.

A key innovation of our work is the characterization of distinct aging types, accelerated and decelerated aging—based on cross-sectional multi-omics data (Fig. \ref{fig:clustering}a,b,c,d). We demonstrate that accelerated aging is not simply a faster version of normative aging but a distinct process marked by an early and large-scale disruption of molecular homeostasis, with the most pronounced dysregulation originating in the microbiome. This disruption manifests as precipitous changes in molecular trajectories during mid-life, typically beginning around age 50. In contrast, decelerated aging is defined by prolonged molecular stability and more gradual, later-life changes. This distinction suggests that a critical window for interventions aimed at promoting healthy aging may lie in mid-life, prior to this sharp inflection point.

Furthermore, we identified distinct, sex-specific waves of molecular change that define these aging trajectories. In the accelerated aging group, females exhibit an intense and sharply-defined wave of change peaking between 50 and 55 years, particularly within the metabolome and lipidome, which strikingly coincides with the average age of the menopausal transition  \ref{fig:functional_analysis}a,b). Males in this group, conversely, display a more protracted and delayed wave peaking 10–15 years later. Crucially, these intense, sex-specific waves were markedly attenuated and desynchronized in the decelerated aging group. This suggests that a key feature of healthy aging is the ability to buffer and mitigate these profound, and likely hormonally-driven, molecular shifts that occur at critical life stages.

From a clinical standpoint, our multi-omics aging clocks demonstrate clear relevance to human health and disease (Fig. \ref{fig:disease}a,b). The biological age scores, particularly from the gut microbiome clock, were significantly associated with multimorbidity in older adults and showed strong, specific associations with the risk of 13 out of 16 common non-cancer diseases, most notably metabolic disorders such as obesity, fatty liver, and hypertension. These associations remained robust after adjusting for chronological age, sex, and BMI, indicating that our clocks capture a dimension of physiological dysregulation and accumulated damage not reflected by traditional risk factors. This positions the multi-omics biological age as a promising surrogate endpoint for clinical trials and a powerful tool for personalized risk stratification and preventive medicine.

The strengths of our study include the large, prospective cohort design, the breadth of our multi-omics profiling, and the application of nonlinear modeling frameworks capable of capturing the complex dynamics of biological systems. However, we also acknowledge several limitations. First, our cohort is composed exclusively of Jewish individuals in Israel, and thus the generalizability of our specific clocks and aging trajectories to other ethnic and geographic populations requires further investigation. Second, while our cross-sectional analyses reveal strong associations with prevalent diseases, the ongoing longitudinal follow-up of this cohort will be critical to validate these clocks as predictors of future disease incidence and mortality. Finally, our study establishes strong associations but cannot definitively prove causation; mechanistic studies are needed to determine whether the observed changes, especially in the microbiome, are drivers or consequences of the aging process.

In conclusion, this study reframes human aging as a dynamic, nonlinear, and deeply heterogeneous process. By integrating multi-omics data, we have developed a robust biological aging clock that captures systemic health and uncovers distinct, sex-specific aging trajectories. With the microbiome emerging as a central indicator of healthspan, our findings provide a powerful framework for quantifying biological age and offer a promising foundation for the development of personalized strategies to monitor health and prevent age-related disease. Future work should focus on validating these findings in diverse populations and leveraging longitudinal data to dissect the causal pathways linking molecular changes to long-term health outcomes.

\section{Methods}\label{sec4}
\subsection{Description of cohort}
The data for this study were derived from a prospective longitudinal cohort of the Human Phenotype Project (HPP), which was launched in Israel in 2018. We included participants from this cohort who completed the baseline visit between February 2019 and January 2023. After data quality control, a total of 10,019 participants were included in the construction and validation analysis of this aging clock. These participants ranged in age from 40 to 70 years old, provided complete clinical phenotype data at the baseline visit, and successfully completed the collection of biological samples for multi-omics analysis.
This study strictly adhered to the ethical principles of the Declaration of Helsinki and was approved by the Institutional Review Board of the Weizmann Institute of Science in Israel (IRB approval number: 1719-1). All participants were fully informed of the study content and signed informed consent before joining the study.
This cohort study is designed to track participants for up to 25 years, with systematic follow-up planned every 2 years. At recruitment, individuals with severe or chronic illnesses were excluded from the study, providing an ideal baseline for studying the normative aging process in a relatively healthy population. The longitudinal nature of this design, with a healthy baseline population, is critical for developing biological age clocks that can capture differences in aging rates and predict future health risks.

To construct a multidimensional multi-omics aging clock, we integrated multiple levels of data collected at baseline from the HPP cohort. The data included a variety of clinical, physiological, behavioral, and multi-omics analysis data, which we divided into 17 groups: multi-omics features and 16 other body systems representing major physiological systems and environmental exposures (Fig. \ref{fig:human-system}).

\subsection{Body system-derived features}
To systematically analyze the interrelationships among physiological modules, we organized the hundreds of measurement parameters and questionnaire information collected at baseline into the following core system-level categories. 

Baseline characteristics (sex, age, and BMI): This category includes the three most basic anthropometric and demographic characteristics: biological sex, chronological age, and BMI. In subsequent association analyses, they were used as core covariates to adjust for more independent relationships between physiological systems.

\begin{itemize}
    \item \textbf{Blood lipids:} Includes routine clinical laboratory test results such as high-density lipoprotein cholesterol (HDL), total cholesterol, non-HDL cholesterol, and triglycerides.
    \item \textbf{Body composition:} Includes 108 fat and lean mass measurements of various body regions (such as limbs and trunk) obtained based on dual-energy X-ray absorptiometry (DXA) imaging. In addition, anthropometric parameters such as weight, height, waist and hip circumference are also covered.
    \item  \textbf{Bone Density:} Includes 182 mineral content measurements of different bone sites based on DXA imaging.
    \item \textbf{Cardiovascular System: }This system assessment integrates multiple test results, including conventional blood pressure measurement, blood pressure ratio calculated by ankle-brachial index (ABI) test, arterial stiffness assessed by pulse wave velocity (PWV), intima-media thickness (IMT) calculated by carotid ultrasound, fundus vascular parameters analyzed by retinal imaging, and cardiac electrical activity captured by resting electrocardiogram (ECG).
    \item \textbf{Insulin Resistance:} Includes 49 blood glucose variability parameters calculated from continuous glucose monitoring (CGM) device monitoring data, as well as laboratory tests of fasting blood glucose and glycosylated hemoglobin (HbA1C) levels.
    \item \textbf{Liver Health: }The evaluation parameters are derived from liver ultrasound and two-dimensional shear wave elastography (2D-SWE) technology, including indicators such as liver viscosity, elasticity, and attenuation. In addition, a series of blood test enzyme indicators related to liver function, such as alkaline phosphatase (ALP) and alanine aminotransferase (ALT), are also included.
    \item \textbf{Renal Function:} Including creatinine, urea, and electrolytes sodium and potassium from blood tests.
    \item \textbf{Hematopoietic System: }Including a series of indicators derived from complete blood counts that can reflect the characteristics and components of red blood cells, such as hemoglobin, hematocrit, and mean corpuscular volume.
    \item \textbf{Immune System: }Including the total number of white blood cells from the complete blood count, as well as the absolute values and percentages of neutrophils, lymphocytes, monocytes, eosinophils, and basophils.

   \item \textbf{Frailty:} Including the results of the hand grip strength test, as well as the limb lean mass data that overlaps with the "body composition" category.
   \item \textbf{Sleep Characteristics:} Data are collected from a home sleep apnea test device that integrates body movement recording, pulse oximetry and sound sensors. Core parameters measuring blood oxygen saturation, respiratory events, heart rate and sleep structure are calculated by dedicated software.

   \item \textbf{Lifestyle and Mental Health:} Data in this section are mainly collected through electronic questionnaires, and its structure and content refer to the standardized questionnaire of the UK Biobank.

   \item \textbf{Lifestyle:} Covers 46 questions such as smoking, drinking, physical activity, employment status and use of electronic devices.

   \item\textbf{Mental Health:} Covers 35 questions on mood, life satisfaction and depressive symptoms, and calculates the recent depressive symptoms (RDS) score based on this.
   \item\textbf{Diet: }The average daily intake of 322 foods is calculated through the participants' real-time self-recording for up to 16 days. The data has undergone strict quality control, such as excluding records with a total daily calorie intake of less than 500 kcal and processing extreme values.
\end{itemize}
\subsection{Transcriptomics}
Transcriptomic profiling was conducted on human peripheral blood mononuclear cells (PBMCs). RNA extraction was performed from PBMC samples obtained from participants at the Clinical Testing Center (CTC) following standard operating procedures. Blood was collected in three tubes per participant (Complete Blood Count, serum, and PBMC isolation). PBMCs were processed within 2 hours of collection by centrifugation to isolate the mononuclear layer. Following isolation, PBMCs were aliquoted into three tubes per sample, and one aliquot designated for RNA sequencing was stored at -80°C to preserve RNA integrity. Cell viability and concentration were evaluated using an automated cell counter to ensure high-quality samples for downstream processing.
For library preparation, total RNA was extracted using a standard protocol. During reverse transcription, unique molecular identifiers (UMIs), sample barcodes, PCR primers, and ERCC spike-ins were incorporated into each sample. ERCC spike-ins served as external controls for quantification and normalization of gene expression levels. The resulting cDNA underwent initial amplification to increase material yield for library construction. Subsequently, a tagmentation reaction was performed using a cutting enzyme (TDE1, or Nextera transposase prior to May 2023) to fragment the cDNA and introduce sequencing adaptors. Following fragmentation, samples were pooled, and a final round of PCR amplification was carried out to complete the cDNA library for sequencing.
RNA sequencing libraries were sequenced on an Illumina NovaSeq platform with an average read depth of approximately 5 million reads per sample. Read 1 (R1) was 20 base pairs in length (containing the UMI and poly-T tail), while Read 2 (R2) was 66 base pairs in length and contained the cDNA fragment and possibly the poly-A tail. PhiX control libraries were spiked into sequencing runs to enhance base diversity. The sequencing generated high-quality 3' RNA-seq data suitable for downstream bioinformatic analyses.

\subsection{Metabolomics}
Metabolite profiling was conducted using the proton nuclear magnetic resonance ($^1$H-NMR) platform of Nightingale Health, for which the technical details and relevant epidemiological applications have been previously described \cite{soininen2015quantitative,wurtz2017quantitative}. This platform provides simultaneous quantification of 228 absolute-value-based plasma metabolites and ratios, with concentrations estimated from reference data. The profiling primarily extends detailed lipidomic characterization and includes measurements of clinically validated biomarkers, such as routine lipids, lipoprotein subclass profiling with lipid concentrations within 14 subclasses, fatty acid composition, and various low-molecular-weight metabolites including amino acids, ketone bodies, and glycolysis metabolites.
For data processing, instances of the value 'TAG' were replaced with zero, and a corresponding binary indicator variable was added. No further normalization or imputation was applied.

\subsection{Serum lipidomics}
Lipid profiling was conducted following a methodology similar to that described by \cite{reicher2024phenome}. Lipid extracts were analyzed using a Waters ACQUITY UPLC system connected to a Vion IMS QTof mass spectrometer (Waters Corp.). 
The separation of lipid species was achieved on an ACQUITY UPLC BEH C8 column (2.1 $\times$ 100 mm, 1.7$\mu$m; Waters Corp.). 
The column temperature was maintained at 40 $^{\circ}$C with a constant flow rate of 0.4 ml min$^{-1}$.
The mobile phase consisted of two solvents: mobile phase A (46:38:16 v/v/v double distilled water:acetonitrile:isopropanol) and mobile phase B (1:69:30 v/v/v double distilled water:acetonitrile:isopropanol). Both phases were supplemented with 1\% 1 M NH$_4$Ac and 0.1\% glacial acetic acid.The total run time for each sample was 10 minutes, employing a linear gradient. The gradient started with $100\%$ mobile phase A for the first 0.5 min. Over the next 3.5 min (until 4.0 min), the proportion of mobile phase A was linearly decreased to $25\%$. It was further reduced to $0\%$ between 4.0 and 6.0 min. Mobile phase B was then maintained at $100\%$ from 6.0 to 8.0 min. The system was rapidly switched back to $100\%$ mobile phase A at 8.1 min and held for re-equilibration until the end of the run at 10.0 min.
Mass spectrometric data were acquired in full-scan HDMSE resolution mode across a mass range of m/z 50-2,000 Da. 
The instrument was operated in both positive and negative ionization modes. The capillary voltage was set to 3.0 kV for positive mode and 2.0 kV for negative mode, with a consistent cone voltage of 40 V. The source temperature was set to 120 $^{\circ}$C and the desolvation temperature to 450 $^{\circ}$C. Nitrogen was utilized as the desolvation gas and cone gas, at flow rates of 800 l h$^{-1}$ and 30 L h$^{-1}$, respectively.
For the high-energy scan function, a collision energy ramp of 20–80 V was applied in positive mode, and 30–80 V in negative mode. A low-energy scan function was performed with a collision energy of 4 V. Leucine-enkephalin was infused as the lock-mass reference to ensure mass accuracy throughout the analysis.
The raw instrumental data files were first exported from the UNIFY software (Waters Corp.) and subsequently imported into Progenesis-IQ using the native Waters format importer plugin. Data from each 96-well plate were processed as a separate block for each ionization mode. Within Progenesis-IQ, adduct annotation was automated, but constrained to detect only doubly charged ions. Following automated processing, all data were manually inspected for quality. Finally, a matrix of compound measurements was exported for further statistical analysis using custom in-house scripts.

\subsection{Microbiome}
Fecal samples were collected using OMNIgene-Gut stool collection kits (DNA Genotek), and subgingival plaque samples were obtained by a dentist using the same type of kit. Genomic DNA was extracted with the PowerMag Soil DNA Isolation Kit (MoBio), optimized for use on a Tecan automated platform. Shotgun metagenomic sequencing was performed on either the Illumina NextSeq 500 (single-end, 75 bp reads) or NovaSeq 6000 (single-end, 100 bp reads) platforms. Raw sequencing reads were processed with Trimmomatic0.38 for adapter removal and quality filtering using the following parameters: -phred33 ILLUMINACLIP:$<$adapter file$>$:2:30:10 SLIDINGWINDOW:6:20 CROP:75 MINLEN:65 for 75 bp reads, and CROP:100 MINLEN:90 for 100 bp reads. Reads mapping to the human genome (hg19) were identified with Bowtie2 and excluded from downstream analyses.
Species-level composition was determined in two steps. First, non-human reads were aligned with Bowtie2 to a subset of representative genomes from Pasolli et al. \cite{pasolli2019extensive}, limited to species-level genome bins (SGBs) containing at least five genomes. Taxonomic assignments were based on a majority vote among reference genomes within each SGB; when reference genomes were absent, higher-level taxonomic labels were assigned following the criteria in Pasolli et al\cite{pasolli2019extensive}. In the second step, species relative abundance was calculated using the Unique Relative Abundance (URA) method, which estimates abundance as the mean coverage of the 50\% most densely covered genomic regions, considering only uniquely mapped reads. Parameters were as follows: for 75 bp gut samples [min\_mapped\_to\_retain = 1 M reads; num\_mapped\_to\_subsample = 8 M reads], for 75 bp oral samples [0.5 M, 4 M reads], for 100 bp gut [1 M, 5 M reads], and for 100 bp oral [0.5 M, 2.5 M reads].
Shannon $\alpha$-diversity and species richness were computed prior to the data processing steps described above. The proportion of human DNA shedding was estimated as the fraction of human reads removed relative to the total number of reads passing quality control.
Microbial functions were calculated using MetaCyc24 pathways by HUMAnN3, Bowtie2, DIAMOND2 and MetaPhlAn4 \cite{caspi2020metacyc,beghini2021integrating,buchfink2021sensitive,blanco2023extending}.

\subsection{Calculation of chronological age measures}

In the HPP dataset, chronological age was not provided as a pre-calculated variable. However, the dataset contains precise dates of birth for each participant, as well as the specific dates for all clinical visits and phenotyping assessments. To support our analyses, we therefore derived a high-precision, decimal-based chronological age for each participant at every relevant time point.
The calculation was performed as follows. For each participant, we identified their date of birth and the date on which a specific assessment (e.g., baseline visit, cognitive testing, or physiological measurement) was conducted. We then calculated the exact number of days between these two dates. To convert this interval into years, the total number of days was divided by 365.25. The use of 365.25 as the divisor accurately accounts for the occurrence of leap years, ensuring a robust and precise age estimation. This procedure was systematically applied to derive the chronological age for every participant at each data collection point used in the subsequent analyses.

\subsection{Feature selection}


To identify a robust and informative subset of age-associated features from the high-dimensional multi-omics data, we implemented a standardized two-stage feature selection strategy. This procedure was designed to first isolate statistically significant features and then to refine this selection based on predictive relevance. Crucially, this strategy was applied independently and consistently across each of the five omics datasets (i.e., transcriptome, metabolome, lipidome, gut microbiome, and oral microbiome) to ensure the methodological uniformity and comparability of all subsequent modeling efforts.
The first stage involved a correlation-based filtering process. For each omics layer, we calculated the Spearman's rank correlation coefficient between every feature and the participants' chronological age. We chose Spearman's method due to its robustness to non-linear relationships and its non-parametric nature, which is well-suited for diverse biological data types that may not follow a normal distribution. The resulting p-values were then adjusted for multiple hypothesis testing using the Benjamini-Hochberg (BH) procedure to control the false discovery rate (FDR). Only features with a BH-adjusted p-value less than 0.05 were retained for the next stage.
In the second stage, the pre-filtered features were subjected to a model-based importance assessment using SHAP\cite{lundberg2020local}. We first trained a LightGBM model, a highly efficient gradient boosting framework, to predict chronological age using the features selected from the first stage. Subsequently, we employed the SHAP framework to compute the contribution of each feature to the model's predictions. SHAP values provide a rigorous, game theory-based measure of feature importance. We selected all features with a mean absolute SHAP value greater than zero, ensuring that only variables with a tangible contribution to the model's predictive output were included in the final feature set. The final sets of features derived from this two-stage procedure for each omics layer were used for all downstream analyses and model construction.

\subsection{Model establishment and evaluation criteria}

To construct robust predictive models of chronological age from the selected omics features, we established and compared a range of machine learning architectures. Our approach was designed to systematically evaluate both linear and non-linear relationships. We implemented two regularized linear models, LASSO and ElasticNet, which are well-suited for high-dimensional data and valued for their interpretability and embedded feature selection. In parallel, we implemented LightGBM, a high-performance gradient boosting framework, selected for its superior ability to capture complex, non-linear interactions that are characteristic of biological systems.
The establishment of all models was performed under a stringent 10-fold cross-validation framework to ensure that performance estimates were robust and generalizable. The entire modeling pipeline, from data splitting to training and prediction, was implemented in Python, primarily utilizing the scikit-learn library. This standardized procedure was uniformly applied to the feature sets derived from each of the five omics platforms (transcriptome, metabolome, lipidome, gut microbiome, and oral microbiome), thereby ensuring a fair and direct comparison of model performance across different data types.
The evaluation criteria for model performance were based on a comprehensive suite of five distinct metrics designed to quantify the concordance between model-predicted age and actual chronological age. These criteria included: (1) the Pearson correlation coefficient (R), to assess the strength of the linear relationship between predicted and actual values; (2) the mean absolute error (m.a.e.), representing the average absolute difference; (3) the root mean squared error (r.m.s.e.), which is sensitive to large errors; (4) the median absolute error (med.a.e), a robust metric against outlier predictions; and (5) the coefficient of determination (R2), indicating the proportion of variance in chronological age explained by the model.
Based on a systematic comparison using these criteria, the LightGBM models consistently outperformed the linear models across all five omics datasets, demonstrating higher accuracy and robustness. Therefore, the optimized LightGBM model for each omics layer was established as the final model and utilized for all subsequent downstream analyses.

\subsection{Pathway enrichment analysis}
We conducted gene ontology (GO) enrichment analysis using the clusterProfiler R package (v4.8.1) to identify biological pathways and functions significantly associated with our gene list. We focused specifically on GO biological processes.
To determine statistical significance, we used the Benjamini-Hochberg method to adjust the P values for multiple testing. Only pathways with an adjusted P-value less than 0.05 were considered significant.
For visualization of the enrichment results, we used the ggplot2 R package (v3.4.1).

\subsection{Trajectory clustering analysis}
To identify distinct molecular trajectory patterns within the combined multi-omics data, we performed a shape-aware time-series clustering analysis. We specifically chose the Fuzzy C-means (FCM) \cite{shen2024multi} algorithm because it assigns each molecule a graded membership to multiple clusters, which better captures partially overlapping, co-regulated trajectories than hard, distance-only partitioning methods.
To prevent features with large variances from dominating the partition, we z-standardized each molecular trajectory across all time points. We then determined the optimal number of clusters k and the fuzzification parameter m using a consensus strategy: the elbow curve of within-cluster dispersion, the average silhouette score across k, and fuzzy-partition validity indices evaluated over a grid of k and m. The k, m pair that jointly optimized these metrics was used for downstream analyses. 
Using the selected k and m, we ran FCM on the standardized trajectories. 

\subsection{DE-SWAN}
The DE-SWAN algorithm \cite{lehallier2019undulating} was used. To begin, a unique age is selected as the center of a 20-year window. Molecule levels in individuals younger than and older than that age are compared using the Wilcoxon test to assess differential expression. P values are calculated for each molecule, indicating the significance of the observed differences.

\subsection{Statistics and reproducibility}
No statistical method was used to predetermine sample size. No data were excluded from the analyses. Data distribution was assumed to be normal, but this was not formally tested.

\section{Data availability}\label{sec5}

Data in this paper are part of the Human Phenotype Project and are accessible to researchers from universities and other research institutions at https://humanphenotypeproject.org/. Interested bona fide researchers should contact info@pheno.ai to obtain instructions for accessing the data.

\section{Code availability}\label{sec6}
Analysis source code is available at https://github.com/LFD-byte/PhenoAI-MOBA.

\bibliography{sn-bibliography}

\end{document}